\documentclass[fleqn,usenatbib]{mnras}

\usepackage{newtxtext,newtxmath}

\usepackage[T1]{fontenc}

\DeclareRobustCommand{\VAN}[3]{#2}
\let\VANthebibliography\thebibliography
\def\thebibliography{\DeclareRobustCommand{\VAN}[3]{##3}\VANthebibliography}

\usepackage{graphicx}	
\usepackage{amsmath}	

\newcommand{\msun}{{\,\rm M_\odot}}

\newcommand{\kms}{\,{\rm km}\,{\rm s}^{-1}}

\def\gsim{ \lower .75ex \hbox{$\sim$} \llap{\raise .27ex \hbox{$>$}} }
\def\lsim{ \lower .75ex \hbox{$\sim$} \llap{\raise .27ex \hbox{$<$}} }

\title[Milky Way halo dark matter decay]{Dark matter decay in the Milky Way halo}

\author[M.~R. Lovell]{
M. R. Lovell$^{1,2}$\thanks{E-mail: m.r.lovell@durham.ac.uk}
\\
$^{1}$ Institute for Computational Cosmology, Durham University, South Road, Durham DH1 3LE, United Kingdom\\
$^{2}$ Department of Physics, Durham University, South Road, Durham DH1 3LE, United Kingdom \\
}

\date{Accepted XXX. Received YYY; in original form ZZZ}

\pubyear{2024}

\begin{document}
\label{firstpage}
\pagerange{\pageref{firstpage}--\pageref{lastpage}}
\maketitle

\begin{abstract}
Dark matter may be detected in X-ray decay, including from the decay of the dark matter particles that make up the Milky Way (MW) halo. We use a range of density profiles to compute X-ray line intensity profiles, with a focus on the resonantly produced sterile neutrino dark matter candidate. Compared to the Navarro--Frenk--White density profile, we show that using an adiabatically contracted halo profile suppresses the line intensity in the halo outskirts and enhances it in the Galactic Centre (GC), although this enhancement is eliminated by the likely presence of a core within 3~kpc. Comparing our results to MW halo observations, other X-ray observations, and structure formation constraints implies a sterile neutrino mixing angle parameter $s_{11}\equiv\sin^{2}(2\theta)\times10^{11}\sim[3,4]$ (particle lifetime $\tau_{28}\equiv\tau/(10^{28}\rmn{sec})\sim[1.0,1.3]$), which is nevertheless is strong tension with some reported non-detections. We make predictions for the likely decay flux that the {\it XRISM} satellite would measure in the GC, plus the Virgo and Perseus clusters, and outline further steps to determine whether the dark matter is indeed resonantly produced sterile neutrinos as detected in X-ray decay.   
\end{abstract}




\section{Introduction}

One of the key mechanisms for discovering dark matter is indirect detection. This method uses rare instances in which dark matter emits electromagnetic radiation -- either through decay or annihilation -- that can be distinguished from astrophysical processes. For example, \citet{Hooper11} observed a gamma-ray signal in the Galactic Centre (GC) consistent with the annihilation of weakly interacting massive particles (WIMPs). However, these measurements are inconsistent with non-detections in dwarf spheroidal (dSph) satellites of our Milky Way (MW) galaxy \citep{McDaniel24}, and it remains to be seen whether the dark matter annihilation origin hypothesis is still viable. 

Another claimed signal is instead at the keV scale. \citet{Bulbul14} reported an excess in X-ray emission at an energy of 3.55~keV in stacks of galaxy clusters, and at the same time \citet{Boyarsky14a} identified a similar feature in the Perseus cluster and in M31. Other targets for which 3.55~keV feature detections have been reported include the GC \citep[][]{Boyarsky15,Hofmann19} and blank sky observations of the MW halo \citet{Neronov16,Cappelluti17}. This line was consistent with the two-body decay of a dark matter particle of mass $7.1$~keV and a decay lifetime $>10^{27}$~sec. Other studies have presented alternative origins for this feature, including charge exchange with sulphur ions \citep{Gu15,Shah16} and a higher background than used elsewhere (\citealp{Dessert20,Dessert24b}, see also \citealp{Abazajian20a,Boyarsky20}). Therefore nature of this signal therefore remains contested.

One key step in discerning the origin of this reported feature is to resolve the line with a very high resolution X-ray spectrometer, both for its presence and for its velocity dispersion. The ability of such a spectrometer to detect any lines present will first establish whether the previously features do indeed originate from a line or were instead part of the background, and furthermore measuring the velocity dispersion in galaxy clusters with large velocity dispersions enables us to discern whether the line is more likely to originate from DM decay ($[400,800]$~$\kms$) or from astrophysical processes ($<200$~$\kms$) \citep{Bulbul14,Lovell19c}. The {\it Hitomi} mission hosted the {\it SXS} spectrometer but failed before it could take sufficient exposures to detect a line in Perseus \citep{HitomiC17}. The successor {\it XRISM} mission launched in September 2023 carrying the {\it Resolve} instrument, which has an energy resolution of $<7$~eV in the energy range 0.3-12~keV. This corresponds to a velocity resolution of $<300$~$\kms$, and is a factor of 10 improvement on the {\it XMM-Newton} resolution ($<80$~eV, $<3400$~$\kms$). The cleanest indication of a dark matter decay origin is to measure the velocity dispersion across multiple spectroscopic bins; however, this width makes the line more difficult to detect than a narrow line of the same total flux. Therefore, it is easier to investigate possible detections first in bright, low velocity dispersion systems and then use the subsequent detection, if any, to motivate deeper exposures on high velocity dispersion targets. Any further evidence will then support proposals to investigate high mass-to-light ratio/low-background galaxies such as the MW's dwarf spheroidal satellites \citep{Lovell15,Jeltema16,Ruchayskiy16}.   

The best available choice for this high luminosity--low velocity dispersion target in the near future is the GC. Its proximity to Earth leads to a very high flux, which may have already been detected \mbox{\citep{Boyarsky15,Hofmann19}}, and the velocity dispersion is capped by the escape velocity of the MW at $\sim200$~$\kms$, although the astrophysical background may be difficult to estimate. The GC is a particularly important target because of its intrinsic connection to the MW halo dark matter decay emission as a whole: the decay signal of the MW halo must be self-consistent from the GC all the way to the opposite side of the sky. There have been several studies of the MW halo, including the GC itself and blank sky observations probing the outer halo. Some have reported detections of a viable 3.55~keV line \citep{Boyarsky15,Cappelluti17,Boyarsky18,Hofmann19} while others have reported upper limits on the possible flux \citep{Carlson15,Sicilian22,Roach23} or even limits so strong that they rule out any dark matter decay at many standard deviations (\citealp{Dessert20,Dessert20b}; see \citealp{Dessert24a} for their blank sky {\it XRISM} predictions.)

One of the key uncertainties in the expected signal from the MW halo is the dark matter density profile. The spherically averaged profile measured from $N$-body simulations of structure formation is well fit by a Navarro--Frenk--White (NFW) density profile \citep{NFW_96,NFW_97}, which is a two parameter fit of the form:

\begin{equation}
    \rho_\rmn{NFW}=\frac{\rho_\rmn{c}}{\left(r/r_\rmn{s}\right)\left(r/r_\rmn{s}+1\right)^2},
\end{equation}

\noindent where $r$ is the distance from the halo centre, $r_\rmn{s}$ is the scale radius and $\rho_\rmn{c}$ is the characteristic density. This profile has an inner cusp of $-1$ and has frequently been used to model the MW halo for X-ray detection purposes \citep{McMillan17,Sicilian22}. These simulations also predict that the halo should be prolate or oblate rather than spherical \citep[e.g.][]{Frenk88}. On the other hand, simulations that model hydrodynamical processes, including gas cooling, feedback from supernova and supermassive black holes, and the redistribution of angular momentum between stars and the dark matter can all have a measurable impact on the halo, with cooling drawing dark matter inwards through adiabatic contraction while feedback may drive it outwards. Different astrophysics prescriptions can lead to very different results \citep{Lovell18b}, but in general contraction leads to steeper halo density profiles around $[5,20]$~kpc whereas feedback evacuates some fraction of the dark matter from within 5~kpc. Evidence for this result arises from two complementary studies: \citet{Cautun20} used GAIA data to infer contraction in the outer parts of the galaxy, while \citet{Portail17} found evidence for a cored profile within 3~kpc with dynamical modelling of the stellar components. Finally, it is also expected that astrophysics makes the centre of the halo more spherical than is found in $N$-body simulations \citep[e.g.][]{Chua19}.

Most prior studies of the MW halo decay signal assume an NFW profile, such as the fit of \citet{McMillan17}: we hereafter refer to this particular profile as Mc17. One exception is \citet{Roach23}, who instead use the contracted MW halo fit of \citet{Cautun20} at $>1$~kpc and a core $<1$~kpc, and found this model slightly relaxed their non-detection limits relative to the NFW model; we label the \citet{Cautun20} profile Ca20. In this paper we build on the contracted halo model by building in the 3~kpc core measured by \citet{Portail17} and placing our results in the context of both anticipated {\it XRISM} results and of structure formation considerations. We compute a series of spherically averaged dark matter density profiles, including NFW profiles and empirical models, compute the flux from these models as a function of observing angle from the GC, compare the results to observations, and show how any future detection must be consistent with results from galaxy clusters with {\it XRISM}. We will not consider the impact of any non-sphericity effects of the halo.

This paper is organised as follows. In Section~\ref{sec:compres} we compute the density and flux profiles, and present the results, and in Section~\ref{sec:conc} we draw our conclusions. 

\section{Computation and results}
\label{sec:compres}

\subsection{Density profiles}

\begin{figure}
    \centering
    \includegraphics[scale=0.34]{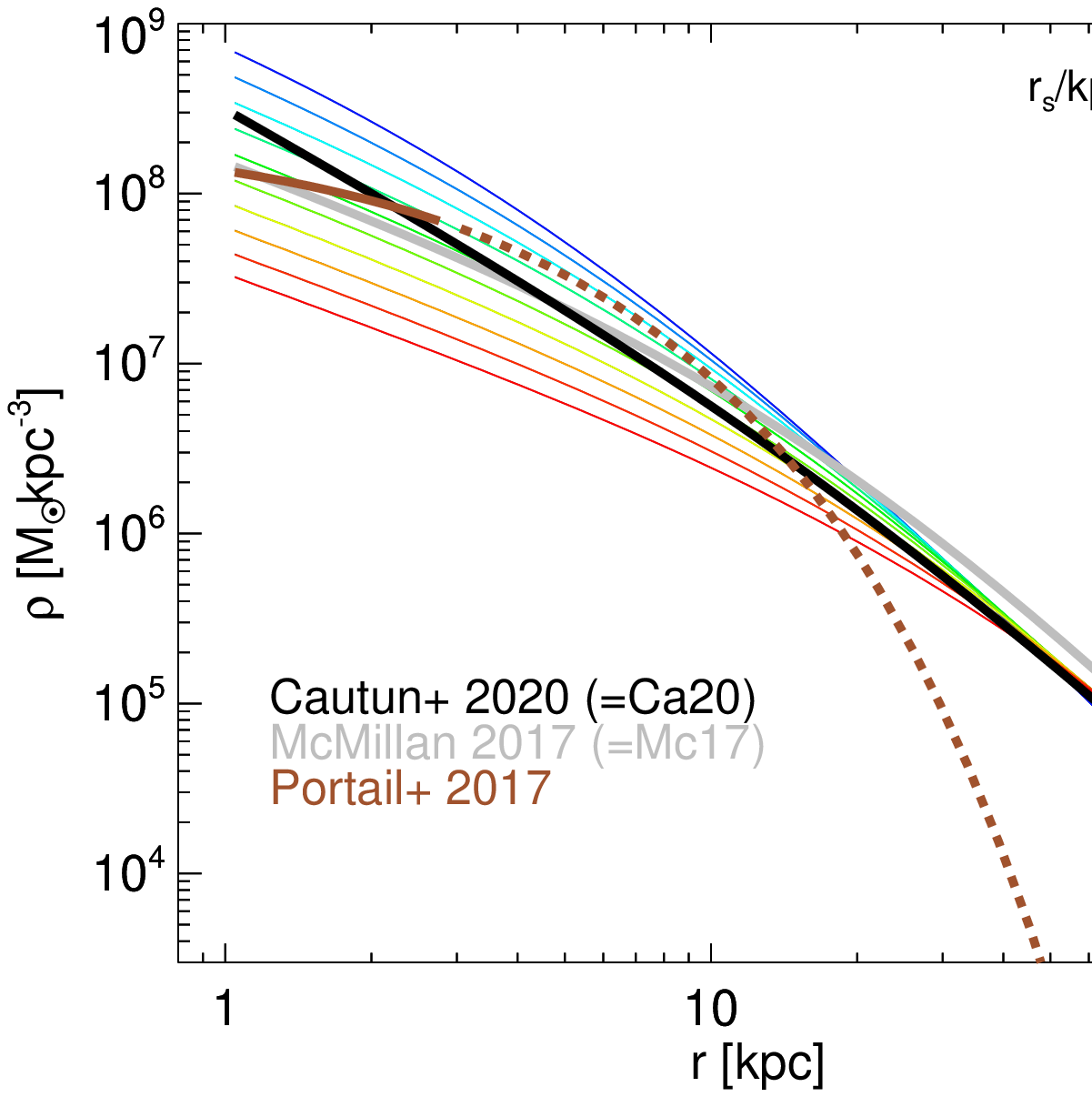}
    \caption{A range of density profiles proposed to model the MW halo. The Ca20 profile is shown as a solid black line and the Mc17 NFW is shown as a grey line. The fit to the central 3~kpc of \citet{Portail17} is indicated with a solid brown line within its fitting range, and continued outwards with a dashed brown line. The series of coloured lines indicates NFW profiles with the same mass as \citet{Cautun20} within 200~kpc but with a range of scale radii as indicated in the legend.}
    \label{fig:densprofs}
\end{figure}

One of the key inputs for the MW halo dark matter decay flux is the amplitude and shape of the dark matter density profile, as discussed in the previous section. We therefore begin our analysis by computing a wide range of density profiles. Our primary profile of interest is the Ca20 empirical, contracted profile. In order to illustrate how the shape of this profile differs from the NFW, we compute 10 NFW profiles that have the same mass enclosed within 200~kpc as Ca20 but cover a wide range in $r_\rmn{s}$, logarithmically spaced between $r_\rmn{s}=5$~kpc and $r_\rmn{s}=50$~kpc. We also include two further profiles from the literature: Mc17 and the fit to the inner 3~kpc by \citet{Portail17}. We display all of these profiles together in Fig.~\ref{fig:densprofs}.   

The contraction of Ca20 leads to important changes with respect to the NFW models. At $20$~kpc its amplitude is approximately the same as the $r_\rmn{s}=23.2$~kpc model but at $3$~kpc it instead has the density of the $r_\rmn{s}=13.9$~kpc model. It matches the Mc17 density at $\sim5$~kpc and drops faster towards the outskirts, leading to a lower overall halo mass. The \citet{Portail17} inner-galaxy fit is much shallower than either model, and intersects Ca20 at just over $2$~kpc from the halo centre. We therefore anticipate that this is the scale where one should transition between the two different sets of fits in order to reflect where each fit best matches the observations. 

\subsection{Line intensity profiles}

Having demonstrated the range of available density profiles, we proceed to compute their expected X-ray line intensity. For this process we adopt four profiles: Ca20, Mc17, and both of these with the \mbox{\citet{Portail17}} core implemented from the halo centre up to the radius at which it intersects the other profile; we will refer to these two combined-fit profiles as Ca20c and Mc17c. We do not attempt to enforce consistency in the enclosed mass as a function of radius: we simply stitch the two density profiles together to form a single density profile, $\rho_\rmn{DM}(r)$, then compute the mass column density along the line of sight, $S_\rmn{DM}=\int\rho_\rmn{DM}dr$.

The second ingredient in computing the flux is the decay rate of the dark matter particles in the X-ray decay channel, which itself is a function of the particle physics parameters. We therefore have a choice of using a generic decay parameter or that of a specific particle physics model. The most popular particle candidate for this decay signal is the resonantly produced sterile neutrino \citep{Dodelson94,Shi99,Abazajian:01b,Dolgov02}, which has the benefit of being part of a larger standard model extension that also explains baryogenesis and neutrino oscillations \citep{Asaka05,Laine08,Boyarsky09a}. The key parameters for computing the decay rate in this model are the dark matter sterile neutrino mass, $M_\rmn{s}$, and the mixing angle with respect to standard model neutrinos, $\sin^{2}(2\theta)$, which, following \citet{Bulbul14,Boyarsky14a,Hofmann19}, give the line intensity:

\begin{equation}
    I_\rmn{DM}=\frac{0.1\rmn{cm^{-2}s^{-1}sr^{-1}}}{3.25}\frac{\sin^{2}(2\theta)}{10^{-11}}\frac{S_\rmn{DM}}{10^{9}\msun\rmn{kpc^{-2}}}\left(\frac{M_\rmn{s}}{7\rmn{keV}}\right)^4.
    \label{eqn:IDM}
\end{equation}

We can then express the decay rate as a particle lifetime, $\tau$, in the following manner:

\begin{equation}
     \tau = 7.2\times10^{29}\rmn{sec}\times\frac{10^{-8}}{\sin^2(2\theta)}\left(\frac{1}{M_\rmn{s}}\right)^5.
\end{equation}

This parameter is useful in comparing to limits that are agnostic about the underlying particle physics model\footnote{Note that the sterile neutrino model has a dominant decay channel into three neutrinos and therefore its true lifetime is shorter than discussed here.}. On the other hand, the mixing angle can be used to compare the X-ray results to sterile neutrino structure formation constraints \citep[e.g. ][]{horiuchi2016,Lovell16,Lovell23b}. We will therefore quote both quantities in the remainder of this paper. Finally, in order to be able to quote parameter values of order $\sim1$, we will define $\tau_{28}\equiv\tau/10^{28}\rmn{sec}$ and $s_{11}\equiv\sin^{2}(2\theta)\times10^{11}$.

\begin{figure}
    \centering
    \includegraphics[scale=0.42]{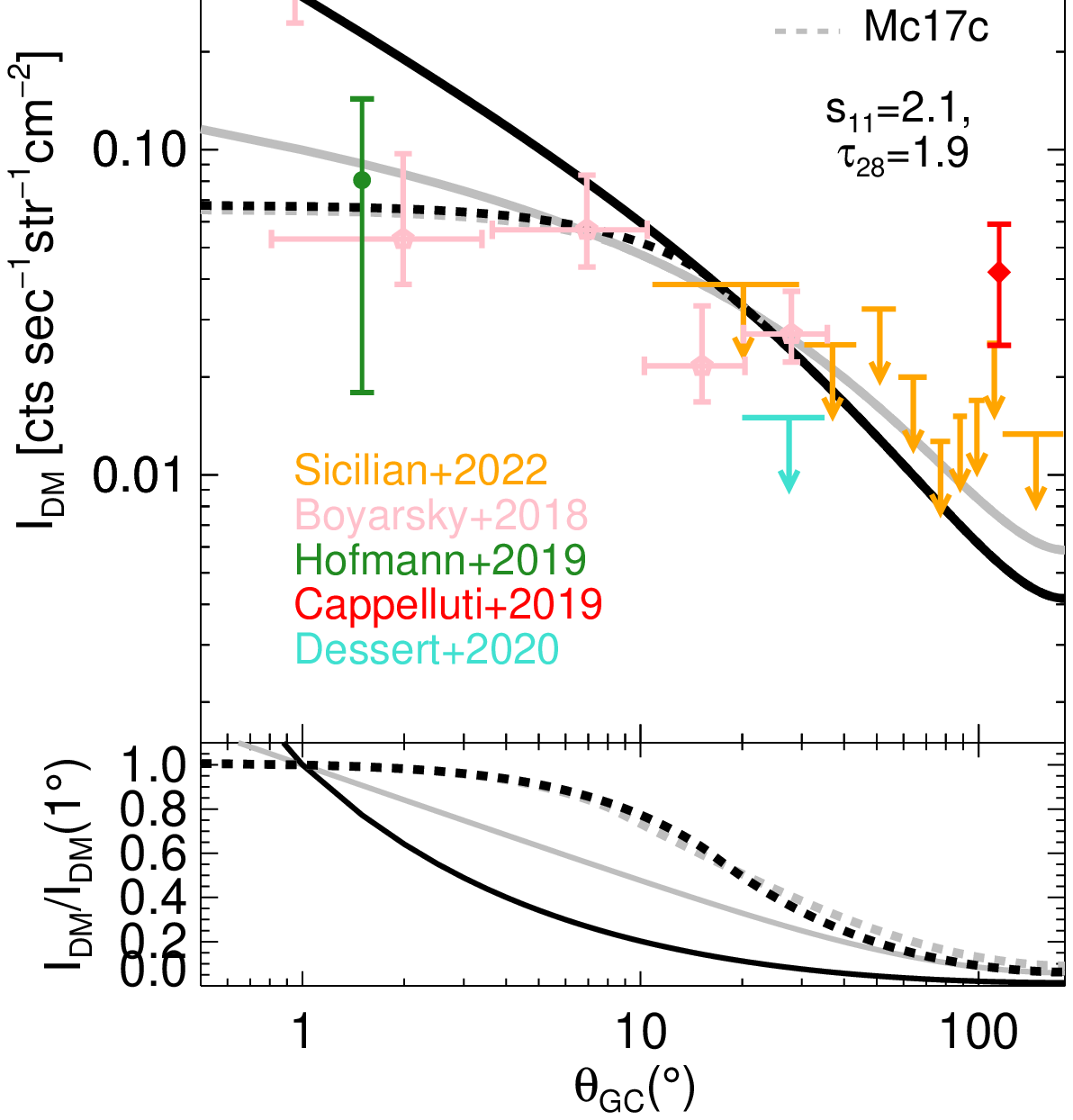}
    \caption{{\it Top panel:} the emission line intensity profile of the MW dark matter decay signal for four density profiles. The \citet{Cautun20}-derived profiles are shown in black and the \citet{McMillan17} equivalents are shown in grey. Profiles modified to include the \citet{Portail17} core (Ca20c, Mc17c) are drawn as dashed lines and those that do not include this core (Ca20, Mc17) are instead drawn as solid lines. The reported detections by \citet{Cappelluti17}, \citet{Boyarsky18} and \citet{Hofmann19} are included as red, pink, and green symbols respectively, and the limits of \citet{Dessert20,Dessert20b} and \citet{Sicilian22}  as cyan and orange symbols. {\it Bottom panel: } the four profiles normalised by their value at $1^{\circ}$ from the GC. }
    \label{fig:intenslog}
\end{figure}

We initially adopt $s_{11}=2.1$ ($\tau_{28}=1.9$) from \citet{Hofmann19} to compute line intensity profiles for the four dark matter density profiles discussed above, and plot the results in Fig.~\ref{fig:intenslog} alongside three reported line detections \citep{Hofmann19,Boyarsky18,Cappelluti17} and two sets of limits \citep{Dessert20,Dessert20b,Sicilian22}. The limits set by \citet{Roach23} are very close to the \citet{Sicilian22}
 limits and are therefore omitted for clarity.

The difference between the Ca20 and Mc17 profiles is significant for GC-angles $\theta_\rmn{GC}<10^{\circ}$, with a factor of three difference between the two at $\theta_\rmn{GC}\sim1^{\circ}$. Conversely, Ca20 is suppressed by 30~per~cent relative to Mc17 at $\theta_\rmn{GC}\sim180^{\circ}$. However, the difference at small angles disappears when the \citet{Portail17} core is included. 

In the same figure we plot the line intensity normalized to its value at $1^{\circ}$. The NFW fit predicts the line intensity to drop by 50~per~cent from $1^{\circ}$ at $10^{\circ}$, compared to $\sim80$~per~cent at the same angle for Ca20 and 20~per~cent for the two cored models. However, the top panel indicates that the smallest degree of uncertainty due to the choice of dark matter profile is at around $20^{\circ}$, where the suppression with respect to the centre is 50~per~cent for the cored models, 60~per~cent for Mc17, and $>90$~per~cent for the Ca20. We therefore conclude that the emission from the GC may vary by up to a a factor of 4 at a fixed $s_{11}$.

We now turn to what dark matter parameters match the observations for the given density profiles. The \citet{Hofmann19} ($\tau_{28}=1.9$) detection preferred a value of $s_{11}=2.1$, and in the same plot we replicate some of the observations discussed in \citet{Sicilian22} and compare them to our set of profiles for this $s_{11}$. The \citet{Sicilian22} results are in reasonable agreement with all of the profiles at $\theta_\rmn{GC}>10^{\circ}$, whereas the \citet{Dessert20,Dessert20b} constraint at the profile-agnostic point of $20^{\circ}$ is in strong tension. Also strongly disfavoured is the reported \citet{Cappelluti17} detection, which is a factor of three higher than Mc17, and therefore is highly unlikely to originate from dark matter decay in the MW halo. The insertion of the \citet{Portail17} core is relevant only for the detections registered within $10^{\circ}$, leading to better agreement with the \citet{Boyarsky18} point at $\sim2^{\circ}$ but worse agreement for their GC point. The \citet{Hofmann19} point is consistent with all curves except for Ca20. 

\begin{figure}
    \centering
    \includegraphics[scale=0.34]{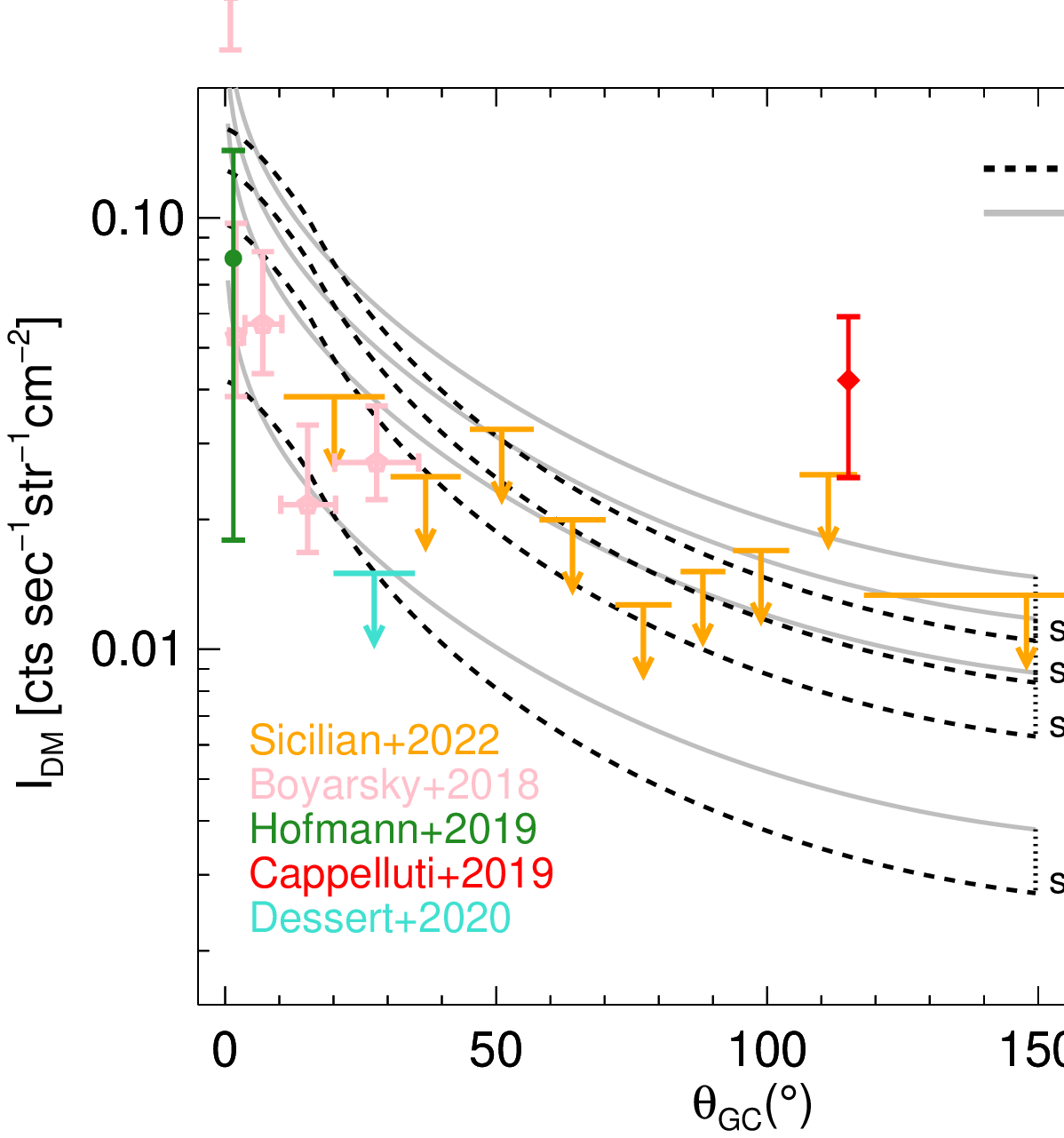}
    \caption{The emission line intensity profile of the MW DM decay flux for a series of four different $s_{11}$ values -- 1.3, 3, 4, 5 -- and two dark density profiles: Ca20c (dashed black lines) and Mc17 (solid grey lines). There is one pair of lines for each $s_{11}$, which are used to label the lines. We also include the observational limits and detections from Fig.~\ref{fig:intenslog} using the same colour scheme.}
    \label{fig:thetas2t}
\end{figure}

We have shown how the four profiles compare to the observations when adopting a specific value of $s_{11}=2.1$ ($\tau_{28}=1.9$). We can adopt further priors on the value of $s_{11}$ with additional X-ray detections and limits, and also from structure formation constraints. This is one of the most remarkable aspects of the resonantly produced sterile neutrino: for the $s_{11}$ of interest, X-ray non-detections constrain high $s_{11}$ and structure formation constrains low $s_{11}$, such that the whole parameter space may be eventually ruled out. A summary of these constraints, including their often contradictory conclusions, is given in fig.~5 of \citet{Lovell23b}: we state here that the  galaxy cluster stack \citep{Bulbul14} and M31 \citep{Boyarsky14a} detections prefer $s_{11}>4$; the galaxy stack of \citet{Anderson14} states $s_{11}<2$, and most structure formation constraints point towards $s_{11}>3$. We explore how these constraints interact the MW halo profile by adopting four values of $s_{11}=[1.3,3,4,5] (\tau_{28}=[3.1,1.3,1.0,0.8])$ and applying these to two of the models: Ca20c and Mc17. We present the results in Fig.~\ref{fig:thetas2t}, using the same plot configuration as Fig.~\ref{fig:intenslog} except for a linear $x$-axis -- in order to emphasise the constraints in the halo outskirts -- and a truncated $y$-axis.  

The $s_{11}=1.3$ model $(\tau_{28}=3.1)$ is the highest value of $s_{11}$ (lowest value of $\tau_{28}$)that is consistent with all of the reported non-detections, including \citet{Dessert20,Dessert20b}, for both of the considered density profiles, but it is incompatible with the reported X-ray detections and with structure formation arguments. At the other end of the chosen $s_{11}$ scale, $s_{11}=5$ $(\tau_{28}=0.8)$ is inconsistent with almost all of the reported non-detections while still predicting intensity profiles significantly lower than the \citet{Cappelluti17} feature. It is also in strong tension with \citet{Hofmann19} result for the Mc17 profile, but much weaker tension for Ca20c due to the core in the latter. The difference between the two profiles is most significant for $s_{11}=3$ $(\tau_{28}=1.3)$, as it is this mixing angle for which Ca20c is consistent with the \citet{Sicilian22} constraints at $\theta_\rmn{GC}>30^{\circ}$ while the shallower Mc17 curve is instead in tension with their results. The two profiles both predict 25~per~cent more line intensity than is consistent with the \citet{Sicilian22} limits at $20^{\circ}$. $s_{11}=4$ $(\tau_{28}=1.0)$ leads to worse agreement at this angle, yet even for this value Ca20c enables agreement with \citet{Sicilian22} at $\theta_\rmn{GC}>50^{\circ}$ whereas Mc17 does not. In conclusion, the combination of these profiles with external constraints prefers $s_{11}\sim[3,4]$, and by extension $\tau_{28}\sim[1.0,1.3]$. 

\subsection{{\it XRISM} flux predictions}

One unavoidable test of any reported dark matter decay detection in the GC is that it must be consistent with detections or non-detections from other targets. Two such objects to be observed by {\it XRISM} in its performance verification (PV) phase are the Virgo and Perseus galaxy clusters. \citet{Lovell23b} argued that the PV exposure of Virgo may be sufficient to return a detection without any further time allocation, whereas the Perseus observation time may be only a quarter of that required. We will therefore make a rough estimate of the expected flux from both of these galaxy clusters, with the goal of first comparing the models to the brighter GC and Virgo targets to the total flux, then motivating further observations to measure the line width in Virgo and make a detection in Perseus, and finally obtaining enough Perseus exposure time to measure its characteristic velocity dispersion of $\sim650$~$\kms$ \citep{Lovell19c}.

\begin{figure}
    \centering
    \includegraphics[scale=0.33]{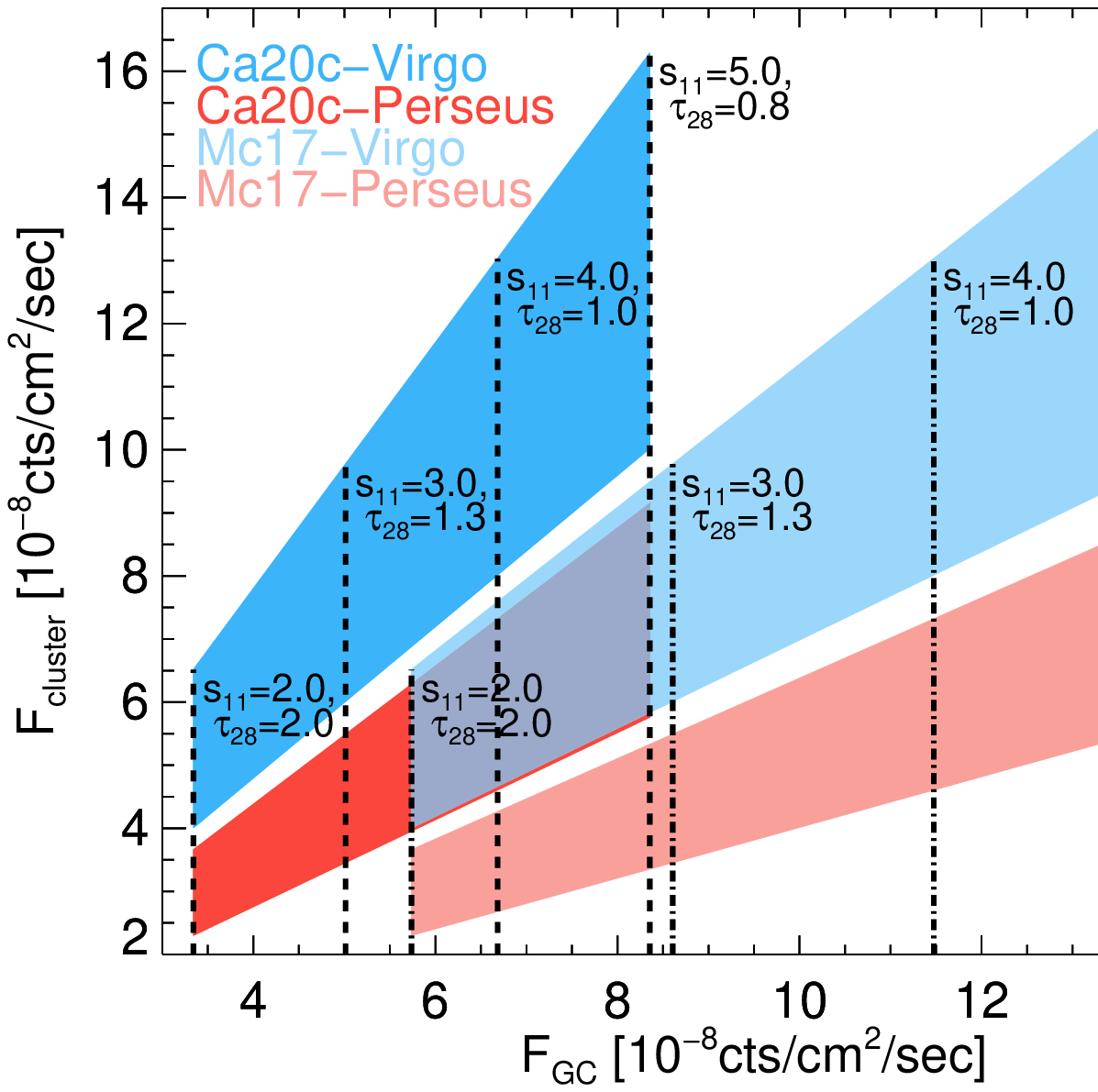}
    \caption{Predictions for the flux to be measured by {\it XRISM} for the GC, Virgo cluster, and Perseus cluster. The flux for the GC is given on the x-axis and for the two clusters on the y-axis. The bright blue and bright red regions indicate the anticipated flux parameter space for GC-Virgo and GC-Perseus pairs respectively when assuming the Ca20c profile for the GC, and the faded regions show the same GC-cluster pairs when instead assuming the Mc17 profile. In both cases we show regions assuming $s_{11}=[2,5]$ ($\tau_{28}=[0.8,2]$), with the specific parameter space for $s_{11}=2,3,4,5$ indicated with vertical lines.}
    \label{fig:GVP}
\end{figure}

We proceed by first computing the GC flux $(\theta_\rmn{GC}=0^{\circ})$ within the {\it XRISM} FoV for $s_{11}=[2,5]$ $\tau_{28}=[0.8,2.0]$ and for the two dark matter profiles considered in Fig.~\ref{fig:thetas2t}. We assume that these two results bracket the range of expected GC flux given the uncertainty in the density profile. We infer counterpart fluxes for the two galaxy clusters by rescaling the results of 500 randomly orientated {\it XRISM} mock observations for analogue cluster hydrodynamical simulations in \citet{Lovell19c} by mixing angle: we use the breadth of the flux distributions across the 500 sightlines to estimate very rough uncertainties on the dark matter distribution. We then compute the range of on-centre Virgo flux and Perseus flux as a function of GC flux, and plot the results in Fig.~\ref{fig:GVP}.

The total range of flux inferred for the GC is $[3,14.5]\times10^{-8}\rmn{cts/cm^{2}/sec}$, while for the preferred $s_{11}=[3,4]$ range ($\tau_{28}=[1.0,1.3]$) it is $[5,11.5]\times10^{-8}\rmn{cts/cm^{2}/sec}$ including the uncertainty in the dark matter density profile. In the same preferred $s_{11}$ range, the Virgo cluster flux is $[6,13]\times10^{-8}\rmn{cts/cm^{2}/sec}$ and the Perseus flux is $[3.5,7.5]\times10^{-8}\rmn{cts/cm^{2}/sec}$. If three emission line signals are detected, they will need to each be consistent with these three flux ranges in order for dark matter decay of resonantly produced sterile neutrino dark matter to be a viable explanation. 

We also note that there will be some contribution to the signals at Virgo and Perseus from the outskirts of the MW halo, which are not included in the flux estimates of Fig.~\ref{fig:GVP}. The Virgo cluster centre is $\sim87^{\circ}$ from the GC, where the MW halo flux contribution is 10~per~cent (8~per~cent) of the GC flux when assuming the Ca20c (Mc17) model; Perseus is at $\theta_\rmn{GC}\sim146^{\circ}$ with MW-halo contributions of 7~per~cent (5~per~cent) of the GC for the Ca20c (Mc17) model. Given that the expected Perseus flux is no less than 35~per~cent of the GC and Virgo is at least 60~per~cent, the MW halo contribution to the cluster-directed signal is $<20$~per~cent for Perseus and $<16$~per~cent for Virgo.   

\section{Conclusions}
\label{sec:conc}

The detection of dark matter annihilation or decay products remains one of the most promising avenues for establishing its particle nature. In the case of the resonantly produced sterile neutrino dark matter candidate, the simplicity of its detection channel -- monochromatic X-ray emission with a resolvable velocity dispersion -- is complemented by structure formation constraints \citep[e.g.][]{Lovell23b}. The recent launch of the {\it XRISM} mission has provided a unique opportunity to test this model, with the prior report of a candidate line at an energy of 3.55~keV \citep{Boyarsky14a,Bulbul14} providing a compelling starting point for decay searches.

In this paper we focus on the MW dark matter halo as a target, due to the anticipated brightness of the GC, the wide range of existing constraints and detections associated with the MW halo \citep{Boyarsky15,Hofmann19,Dessert20,Dessert20b,Sicilian22,Roach23}, and the uncertainty in the MW dark matter density profile. We compute a series of spherically averaged density profiles, including an NFW fit (Mc17), an empirical fit that includes adiabatic contraction (Ca20), and a cored fit to the inner 3~kpc (Fig.~\ref{fig:densprofs}). We show that the contracted Ca20 profile strongly enhances the flux intensity profile at $\theta_\rmn{GC}<20^{\circ}$ relative to the Mc17 NFW, and by contrast suppresses it at larger angles (Fig.~\ref{fig:intenslog}). The introduction of a \citet{Portail17} core instead lowers the intensity within $\theta_\rmn{GC}<10^{\circ}$. 

We explore the impact of changing the mixing angle $s_{11}$, and thus the measured particle lifetime $\tau_{28}$, alongside the dark matter profile with respect to the observations. The switch from Mc17 to Ca20 enables some agreement with observations for $s_{11}=4$ ($\tau_{28}=1.0$) at $\theta_\rmn{GC}>50^{\circ}$, while the core applied in the Mc17c and Ca20c profiles is crucial to preserving the agreement with the \citet{Hofmann19} GC detection (Fig.~\ref{fig:thetas2t}). When combining the results with detections and constraints from other sources \citep{Lovell23b}, we find a rough preference for $s_{11}\sim[3,4]$ ($\tau_{28}\sim[1.0,1.3]$). The greatest challenge to this interpretation is at $\theta_\rmn{GC}\sim25^{\circ}$, where the dark matter curves are in some tension with the \citet{Sicilian22} constraints and are not compatible with \citet{Dessert20,Dessert20b}.

We compute the expected flux to be measured by {\it XRISM} from the GC for the Mc17 and Ca20c, and quoted these results alongside inferred fluxes for on-centre observations of the Virgo and Perseus clusters (Fig.~\ref{fig:GVP}). Using the two profiles to bookend the possible flux and assuming $s_{11}=[3,4]$, we predict a GC flux of $[5,11.5]\times10^{-8}\rmn{cts/cm^{2}/sec}$, a Virgo cluster flux of $[6,13]\times10^{-8}\rmn{cts/cm^{2}/sec}$ and a Perseus cluster flux of $[3.5,7.5]\times10^{-8}\rmn{cts/cm^{2}/sec}$. All of these values must be measured in order to support the dark matter decay hypothesis.

The {\it XRISM} PV observations may be sufficient to detect these fluxes in the GC and Virgo but not in Perseus, although measuring the line velocity dispersions in the clusters would provide an especially strong piece of evidence in favour of a dark matter decay origin. Important follow up observations would include detections in low background targets such as the Draco dwarf galaxy \citep{Lovell15} (see Appendix~\ref{app:UMIII} for a discussion of the recently discovered Ursa Major III/UNIONS 1 object) and the Bullet cluster \citep{Boyarsky08a}, and detection of the MW halo line centroid shift induced by the Earth's motion \citep{Speckhard16}. We therefore anticipate the following work flow for X-ray analysis, where each step must return a positive detection in order to proceed to the next step. We do not attempt to compute the {\it XRISM} time required to make any of these detections, instead we quote the PV time allocations alongside a rough expectation of the order in which targets would become ripe for detection efforts.

\begin{itemize}
    \item PV: measure GC flux $[5,11.5]\times10^{-8}\rmn{cts/cm^{2}/sec}$ (200~ksec Priority A plus 100~ksec in Priority C) and Virgo cluster flux $[5,11.5]\times10^{-8}\rmn{cts/cm^{2}/sec}$ (500~ksec in Priority A).
    \item Extended PV: measure Virgo cluster velocity dispersion $\sim450$~$\kms$ and Perseus cluster flux $[3.5,7.5]\times10^{-8}\rmn{cts/cm^{2}/sec}$  (280~ksec in priority A, plus 100~ksec in priority C).
    \item Further {\it XRISM} observations: observe Perseus velocity dispersion $\sim650$~$\kms$.
    \item Further general instrument observations: observe Draco dSph, Bullet cluster. 
\end{itemize}

The final stage of using the Draco dSph and the Bullet cluster may be well suited to the larger FoV associated with the NewATHENA and proposed Lynx missions. Coupled to attempts to better measure the astrophysical backgrounds we therefore anticipate an excellent opportunity to test the resonantly produced sterile neutrino as dark matter within the next 15 years. 

\section*{Acknowledgements}

MRL would like to thank Oleg Ruchayskyi for comments on the text and Chris Done for useful conversations.

\section*{Data Availability}

Researchers interested in using these results should contact the corresponding author.



\bibliographystyle{mnras}




\appendix
\section{Ursa Major III/UNIONS 1}
\label{app:UMIII}

The recently discovered Ursa~Major~III/UNIONS~1 object \citep{Smith24} would, if confirmed as a dwarf galaxy, be one of the closest and highest density satellites known \citep{Errani24}. Its high density makes it a compelling target for dark matter annihilation studies \citep{Crnogorcevic24,Errani24}, given that the annihilation signal scales as the density squared. By contrast, the decay signal is only sensitive to the mass within the field-of-view (eqn.~\ref{eqn:IDM}). \citet{Errani24} determine that its total mass is $10^{4}$~$\msun$; if we assume that the mass fits entirely within the {\it XRISM-Resolve} field-of-view, the expected flux is $1.3\times10^{-8}(s_{11}/4)$~$\rmn{cts/cm^{2}/sec}$.


\bsp	
\label{lastpage}
\end{document}